\documentstyle[psfig]{elsart}
\newcommand{\req}[1]{(\ref{#1})}
\newcommand{\be}{\begin{equation}}
\newcommand{\ee}{\end{equation}}
\newcommand{\bea}{\begin{eqnarray}}
\newcommand{\eea}{\end{eqnarray}}
\newcommand{\beas}{\begin{eqnarray*}}
\newcommand{\eeas}{\end{eqnarray*}}

\newcommand{\pr}[1]{\left(#1\right)}
\newcommand{\cro}[1]{\left[#1\right]}

\newcommand{\avg}[1]{\langle{#1}\rangle}
\newcommand{\Avg}[1]{\Big<{#1}\Big>}

\newcommand{\erf}{\textrm{erf}}

\newcommand{\Tr}{\textrm{Tr}}

\begin{document}
\title{Colored minority games}
\author{Matteo Marsili$^{(1)}$ and Maurizio Piai$^{(2)}$} 
\address{$^{(1)}$ Istituto Nazionale per la Fisica della
Materia (INFM), Trieste-SISSA Unit, V. Beirut 2-4, Trieste I-34014\\
$^{(2)}$ SISSA-ISAS, V. Beirut 2-4, Trieste I-3401 and Istituto
Nazionale di Fisica Nucleare (INFN),sez. Trieste} \date{\today}

\maketitle

\begin{abstract}
We study the behavior of simple models for financial markets with
widely spread frequency either in the trading activity of agents or in
the occurrence of basic events. The generic picture of a phase
transition between information efficient and inefficient markets still
persists even when agents to trade on widely spread time-scales. We
derive analytically the dependence of the critical threshold on the
distribution of time-scales. We also address the issue of market
efficiency as a function of frequency. In an inefficient market we
find that the size of arbitrage opportunities is inversely
proportional to the frequency of the events on which they
occur. Greatest asymmetries in market outcomes are concentrated on the
most rare events. The practical limits of the applications of these
ideas to real markets are discussed in a specific example.
\end{abstract}

\section{Introduction}

The Minority Game (MG) \cite{CZ1} provides an ideal playground for
studying interacting agent systems such as financial
markets~\cite{MMM}. Its stationary states have indeed been
understood in great detail \cite{CMZ,MCZ,errata} whose
conclusions have been recently confirmed by a fully dynamic
theoretical approach \cite{Coolen}. Such a state of the art allows
one to address interesting issues on the way in which such complex
systems behave. 

The MG is a largely oversimplified model of a systems of adaptive
agents. This characteristic is what makes the MG analytically
tractable thus allowing one to derive a coherent picture of its
collective behavior. Then, it is important to understand how this
picture changes as one moves towards more realistic cases. Several
investigations have indeed aimed at relaxing the many unrealistic
features of the MG: The effect of non-adaptive agents (also called
producers or hedgers) has been studied in Ref.~\cite{MMM}; the role of
market impact was addressed in Refs. \cite{CMZ,MC01,MMRZ} and
differently weighted agents were discussed in Ref. \cite{CCMZ}. Agents
who can refrain from playing, if that is not convenient have also been
considered in Ref. \cite{Johns,BouchMG,CMZ01}. The analytic theory
offers insights on these issues which allows one to form a far more
complete picture than that resulting only from numerical simulations
\cite{web}.

An unrealistic feature of the MG is that agents trade on the same
time-scales and that different events also occur with the same
frequency. Our aim in this note is to study how the collective
behavior changes if one goes beyond such simplifying assumptions.

The MG is based on a {\em syncronous} dynamics. Every agent, at each
time trades in the market. In the real world agents trade on different
frequencies; some everyday, even several times in a day, and some once
a week, a month or an year. The distribution of trading time-scales
across agents may likely be spread over several decades, which may
offer an explaination for the presence of long range correlations in
the volume of financial activity \cite{BouchMG}. How does the behavior
of MG changes if we account for strategies acting on different, even
widely spread, frequencies? 

We find that the qualitative behavior of the system is not
affected. Still a phase transition between an informationally
efficient phase and an inefficient phase exist. We quantify how the
critical point where the phase transition occurs depends on the
distribution. 

The second question we ask concerns the frequency with which events
occur. In the world of the MG one of many possible events occur in
each period. Agents observe these events and indeed they adopt
strategies which provide actions conditional to events. All events in
the MG occur with the same probability. In section \ref{Events} we
study a situation where events occur with different probabilities. Our
discussion builds up on results derived in \cite{relevance_mem}. In
particular we address the issue of information efficiency.

When the market is not informationally efficient, one may investigate
market predictability as a function of the frequency of events. We
find that markets are more predictable when rare events occur than
when typical events occur. In particular the asymmetry of market
outcome in a given state is inversely proportional to the frequency of
that state. This gives a frequency-dependent characterization of market
(in)efficiency.

When comparing these results with real market data, one faces the
problem of identifying relevant events. We discuss a simple example
based on FX data. Average returns of the FX rate are conditioned to
events which, as in the the original MG, encode the information on the
$M$ most recent market movements. Our evidence is not strong but it
points in the right direction: The market is more predictable when
rare events occur.

\section{The Minority Game}

The MG depicts a market with $N$ adaptive agents or speculators.  The
market can be in one of $P$ states, labelled by $\mu$ in the
following. Each trader $i$ has two personal trading strategies,
labeled by a spin variable $s_i$, which prescribe an action
$a^{\mu}_{s_i,i}$ for each state $\mu$. These trading strategies
are drawn at random from a certain distribution (see later) and
assigned to agents. 

The game is repeated many times; at time $t$ the state $\mu(t)$ is
drawn from a distribution $\rho^{\mu}$ at each time and agents try to
estimate, on the basis of past observations, which of their strategies
is the best one. To do this, agents assign a {\em score} $U_{s,i}(t)$
to each strategy $s=\pm 1$. The larger the score of a strategy, the
more likely it will be played by the agent: If $s_i(t)$ is the
strategy played by agent $i$ at time $t$ we assume that
\begin{equation}
{\rm Prob}[s_i(t)=s] \propto \exp\left[ \Gamma U_{s,i}(t) \right]
\label{eq:probability_definition}
\end{equation}
where $\Gamma$ is a {\em learning rate}. Each agent monitors
the scores $U_{s,i}(t)$ of each of her strategies $s$ by
\begin{equation}
U_{s,i}(t+1) = U_{s,i}(t) - a^{\mu(t)}_{s,i} A(t)/N,
~~~A(t)=\sum_j a^{\mu(t)}_{s_j(t),j}.
\label{eq:evolution}
\end{equation}
This dynamics implies that agents prefer to take actions with the
opposite sign to that of $A(t)$. Indeed they reward by increasing
$U_{s,i}$ the score of a strategy $s$ for which
$a_{s,i}^{\mu(t)} A(t)<0$. 
If only binary actions
$a^\mu_{s,i}=\pm 1$ are allowed, then $A(t)$ has the sign of the
action taken by the majority. Hence the minority ``wins''. We refer
the interested reader to Refs. \cite{CZ1,CCMZ,Praha} for deeper
discussions on the interpretation of this dynamics. 

The relevant parameter~\cite{Savit} is the ratio 
\[
\alpha=P/N
\] 
between the ``information complexity'' $P$ and the number of agents,
and the key quantity we shall look at is 
\[
\sigma^2=\avg{A^2}
\]
where $\avg{\ldots}$ is a time average in the stationary
state. $\sigma^2$ is a measure of the inefficiency of agents'
coordination because. Hence optimal states correspond to
minima of $\sigma^2$.

A further interesting quantity is the predictability
\be
H=\sum_{\mu=1}^P \rho^\mu\avg{A|\mu}^2
\label{H}
\ee
\noindent
of the market. Here $\avg{\ldots|\mu}$ is a conditional average in the
state $\mu$. If the average $\avg{A|\mu}$ of $A(t)$ conditional to
$\mu(t)=\mu$ is non-zero, then the sign of $A(t)$ is statistically
predictable if $\mu$ is known. This is why $H$ measures
predictability. 

Previous works have mainly focused on strategies drawn from the
distribution
\be
P(a_{s,i}^\mu)=\frac{1}{2}\delta(a_{s,i}^\mu+1)+
\frac{1}{2}\delta(a_{s,i}^\mu-1)
\label{Pasimu}
\ee
(a correlation between the two trading strategies has also been
considered\cite{MMM}) and on the event distribution
\be
\rho^\mu=\frac{1}{P}.
\label{rhomu}
\ee

The collective behavior of the market is the following: For
$\alpha>\alpha_c$ the market is predictable ($H>0$), the stationary
state is unique and independent of the learning rate $\Gamma$. As
$\alpha$ decreases from large values, also $\sigma^2$ decreases
signalling that agents manage to coordinate more and more
efficiently. At $\alpha=\alpha_c$ a phase transition occurs. Indeed
for $\alpha<\alpha_c$ the market is unpredictable ($H=0$), the
stationary state is not unique but rather depends both on the initial
conditions and on the learning rate. For symmetric initial conditions
($U_{s,i}(0)=0$ $\forall s,i$), $\sigma^2$ reaches a minimum at
$\alpha_c$ and then it starts raising as $\alpha$ decreases below
$\alpha_c$. In this region $\sigma^2$ is also an increasing function
of $\Gamma$. The inefficiency $\sigma^2$ also decreases as a function
of the asymmetry of initial conditions. For strongly asymmetric
initial conditions (e.g. $U_{+,i}(0)\gg U_{-,i}(0)$, $\forall i$)
$\sigma^2$ is greatly reduced and it decreases with $\alpha$.

Such a complex behavior can be captured and described quantitatively
by a statistical mechanics approach \cite{errata} based on two steps:
First one studies the stationary state properties of the system and
secondly one studies the stochastic dynamics. Here we discuss how this
picture changes when either of Eqs. (\ref{Pasimu},\ref{rhomu}) is
modified to allow for agents acting or events occurring on different
frequencies. 

\section{Agents trading on different time-scales}
\label{minor}

Let us assume that agent $i$ plays with a frequency $f_i$. More
precisely we assume that 

\be
P(a_{s,i}^\mu)=f_i\left[\frac{1}{2}\delta(a_{s,i}^\mu+1)+
\frac{1}{2}\delta(a_{s,i}^\mu-1)\right]+(1-f_i)\delta(a_{s,i}^\mu).
\label{Pasimunew}
\ee 
\noindent
This means that on a fraction $1-f_i$ of the events, agent $i$ does
not trade ($a_{s,i}^\mu=0$). Different agents play with different
frequencies, and are not supposed to respect any rigid time-table.  We
assume $f_i$ to be fixed at the beginning of the game, sorting it
randomly on the basis of a given distribution $\nu(f)$. In order to
describe the lack of a single characteristic time-scale, we will
mainly consider the distribution 

\be 
\nu(f)=\gamma\,f^{\gamma-1},
\label{nuf}
\ee 

\noindent
so that the system is now described by two
parameters $\gamma$ and $\alpha$. In the limit $\gamma
\rightarrow \infty$ we recover the case in which all players are
supposed to act at each run of the game.  
Following Ref. \cite{errata}, we find that the stationary state
``magnetizations'' $m_i=\avg{s_i}$ are coincide with the minima of $H$
where, in Eq. \ref{H},

\[
\avg{A|\mu}=\sum_{i=1}^N \left[\frac{a_{+,i}^\mu+a_{-,i}^\mu}{2}
+m_i\frac{a_{+,i}^\mu-a_{-,i}^\mu}{2}\right].
\] 

\noindent
The properties of the stationary states are then accessible using
standard methods of statistical mechanics of disordered systems. The
approach follows exactly the same steps as those of
Refs. \cite{CMZ,MMM}. As detailed in the appendix, the main difference
concerns the disorder average of the replicated partition function
that is now taken with Eq. \req{Pasimunew} and not with
Eq. \req{Pasimu}. Furthermore one has to average over the distribution
$\nu(f)$. Details of the calculations are reported in the appendix.

\begin{figure}
\centerline{\psfig{file=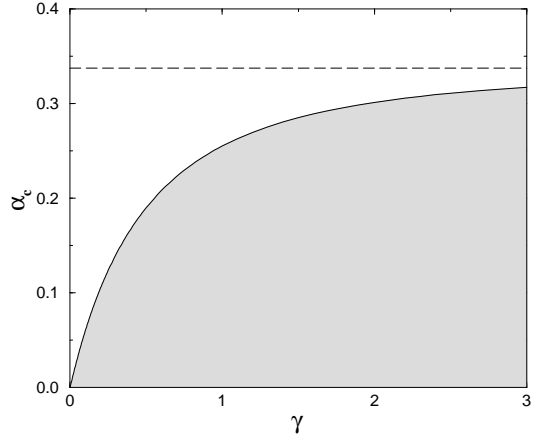,width=7cm}}
\caption{Phase diagram in the $(\alpha,\gamma)$ plane. The shaded
region below the curve $\alpha_c(\gamma)$ is the efficient (symmetric)
phase. The dashed line corresponds to
$\alpha_c(\infty)\approx0.3374\ldots$}
\label{alphacritico}
\end{figure}

The system behaves in a non-trivial way, presenting a phase transition
in the $(\gamma,\alpha)$ plane as shown in
figure~\ref{alphacritico}). This separates an informationally
inefficient phase with $H>0$ -- for $\alpha>\alpha_c(\gamma)$ -- from
an efficient phase ($H=0$) for $\alpha<\alpha_c(\gamma)$. 
As expected when $\gamma\to\infty$ we
recover the value of the critical threshold $\alpha_c\approx
0.3374\ldots$ of the mono-chromatic MG \cite{CMZ}. For finite values
of $\alpha$ we find a smaller value of $\alpha_c$. This is consistent
with the observation that a smaller number of agents are active in the
market at each time.

When $\gamma\to 0$ the critical threshold $\alpha_c(\gamma)$
vanishes. This is consistent with the observation that in such a
market the effective number of trader is a vanishing fraction of $N$.

Concerning agent's behavior we find that the distribution of
magnetization $m$ conditional to the frequency $f$ is given by: 

\be
P(m|f)=\frac{1}{2}\phi(f)
\left[\delta(m+1)+\delta(m-1)\right]+
\frac{\zeta f}{\sqrt{2\pi}}e^{-(\zeta f m)^2/2}
\label{Pmf}
\ee

\noindent
where $\zeta=\zeta(\alpha,\gamma)$ is a parameter which depends on
$\alpha$ and $\gamma$ (see appendix) and 
\[
\phi(f)={\rm erfc}\left(\frac{\zeta f}{\sqrt{2}}\right).
\]
is the probability that an agent with $f_i=f$ is {\em frozen}, i.e.
that he/she plays always the same strategy ($m_i=\pm 1$).  Hence
agents who play less frequently (small $f_i$) are more likely to be
frozen. This is reasonable because it is the market impact \cite{MCZ}
which makes agents switch between strategies and agents who trade less
frequently are less affected by the market impact. 

Most of the {\em noise traders} -- i.e. of the agents with $m_i\approx
0$ -- are frequent traders: It is not difficult to see that the
probability that a trader with $m_i\approx 0$ has frequency $f_i=f$ is 
${\rm Prob}(f_i=f|m_i\approx 0)\propto f^\gamma$. 

The main conclusion of this section is that, even with a broad
distribution of trading time-scales across agents, the basic picture
of the behavior of the MG remains that of the mono-chromatic MG. 
We also see that frequent traders are responsible for market
volatility. 

\section{Broadly distributed frequency of events}
\label{Events}

In this section we address the behavior of a market, as described by
the MG, when the basic events have a broad distribution of
frequencies. In other words, instead of Eq. (\ref{rhomu}) we consider 

\be
\rho^\mu=\frac{1}{P}\tau^\mu
\label{rhomunew}
\ee
\noindent
with $\tau^\mu$ distributed with a pdf $\theta(\tau)$ such that
\[
\int_0^\infty \theta(\tau)\tau d\tau=1
\]
which ensures normalization of $\rho^\mu$ in Eq. \req{rhomunew}.
Now there will be events which occur more frequently than others
with a spread of frequencies which depends on the distribution
$\theta(\tau)$. 

The solution of the MG with a generic $\rho^\mu$ has been
discussed in detail in Ref. \cite{relevance_mem}. Here we pick the
main results: 

The predictability can be written as
\[
H=\frac{1}{P}\sum_{\mu=1}^P \tau^\mu \avg{A|\mu}^2
\cong \int_0^\infty \!d\tau 
\theta(\tau) \tau \avg{A|\tau}^2
\]
where $\avg{A|\tau}^2$ is the average predictability of events with
frequency $\tau$. This quantity can be read off from Eq. (C2) of
Ref. \cite{relevance_mem}, which reads:

\[
\frac{H}{N}=\frac{1+Q}{2}\int_0^\infty \!d\tau 
\theta(\tau)\frac{\tau}{(1+\tau\chi)^2}
\]
where $\chi$ and $Q$ are given in appendix \ref{appAmu}. From these
equations we find that the typical asymmetry $|\avg{A|\tau}|$ that one
can expect on events with frequency $\tau$ is

\be
|\avg{A|\tau}|=\sqrt{\frac{1+Q}{2}}\frac{1}{1+\tau\chi}.
\label{Atau}
\ee

The quantity $|\avg{A|\tau}|$ measures the maximal excess return that
a trader can possibly exploit by trading on events at time-scale
$\tau$.  In a market where agents behave randomly ($m_i=0$ for all
$i$) we expect $|\avg{A|\tau}|$ to be independent of
$\tau$. Eq. \req{Atau} shows that traders activity exploits more
heavily arbitrages on frequent events thus reducing
$|\avg{A|\tau}|$. The result of this is that, for frequent events
($\tau\chi\gg 1$), the excess return is inversely proportional the
probability $\tau/P$ of events. 

Because the number of traders is finite and their ability of detecting
and exploiting arbitrages is bounded, rare events ($\tau\chi\approx
1$) carry arbitrages which are left unexploited. There is a
characteristic frequency $1/\chi$ below which the law
$|\avg{A|\tau}|\sim 1/\tau$ levels off: For rare events
$\rho^\mu<1/(\chi P)$ arbitrage size remains constant, suggesting that
traders are unable to optimize their behavior at a time resolution
smaller than $1/\chi$.

In the limit $\alpha\to\alpha_c^+$ the market becomes efficient so
$|\avg{A|\tau}|\to 0$. At the same time, however $\chi\to\infty$ which
means that the $|\avg{A|\tau}|\sim 1/\tau$ behavior extends to
extremely rare events. 

\begin{figure}
\centerline{\psfig{file=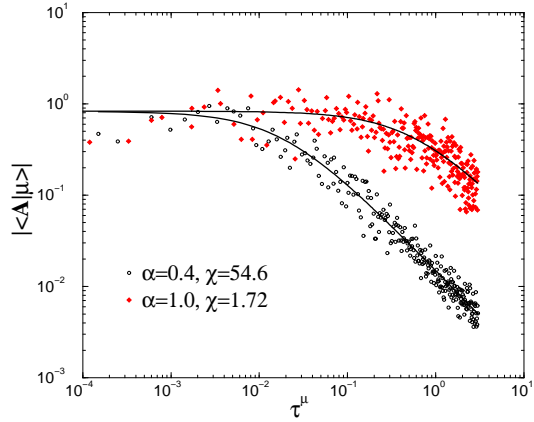,width=7cm}}
\caption{Excess return $|\avg{A|\mu}|$ as a function of $\tau^\mu$ from
numerical simulations of sistems with $P=256$ events, $\alpha=0.4$
and $1.0$. The simulation used $\theta(\tau)=a\tau^{-1/3}$ for
$\tau<\tau_>$. The full lines are numerical fits with the functional
form $A/(1+\chi\tau)$.}
\label{figAmu}
\end{figure}

Fig. \ref{figAmu} shows that numerical simulations are in full
agreement with Eq. \req{Atau}. This plot in particular allows one to
measure $\chi$ from market data. This in turn provides a measure of
the inefficiency of the market. 

\subsection{Efficiency in real markets}

Let us discuss a practical application to the above results to real
market data. There are two main sources of problems: First it is not
clear how to identify a set of exclusive events. Events are specified
in the very definition of the MG but it is not clear what are their
counterparts in a real market. 

Second one faces the problem of handling a finite data set. Suppose
that we have $T$ data points and that, having identified a set of
events, we define $\mu(t)$ as the label of the event which occurs in
observation $t=1,\ldots,T$. Then our estimate $\hat{\avg{A|\mu}}$ of
the market return conditional to event $\mu$ will be
\[
\hat{\avg{A|\mu}}=\frac{1}{T^\mu}\sum_{t=1}^T\delta_{\mu,\mu(t)}A(t),
~~~~T^\mu=\sum_{t=1}^T\delta_{\mu,\mu(t)}.
\]
This will however be affected by statistical errors of the order
\[
\delta\hat{\avg{A|\mu}}\simeq\sqrt{V^\mu/T^\mu},
\]
where $V^\mu$ is the sample variance, conditional to $\mu(t)=\mu$.
One should then require that $\hat{\avg{A|\mu}}\gg
\delta\hat{\avg{A|\mu}}$, i.e. that the results are statistically
significant. This means that market data should be classified in a
number of events as small as possible. However grouping different
market conditions in the same state $\mu$ may average out relevant
information. 

\begin{figure}
\centerline{\psfig{file=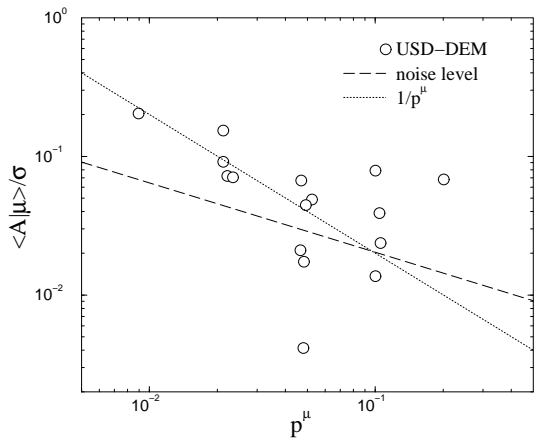,width=7cm}}
\caption{Conditional excess returns $|\avg{A|\mu}|$ as a function of
the frequancy $f^\mu$ in FX markets. For a definition of events
$E^\mu$ see Refs. [17,18].}
\label{AmuFX}
\end{figure}

We illustrate these points with a specific example. We analyzed a data
set of FX data on the USD-DEM exchange rate for the year 1993
\cite{Galluccio}. Considering a typical time interval of 5 minutes,
this gives $T\approx 10^5$ points. We define event $\mu$, as in the
original MG, as the encoding of the $M$ most recent signs of the FX
rate increments \cite{notamu}. Taking $M=4$, i.e. $P=16$ events, we
find the results of Fig. \ref{AmuFX}. The error on
$\hat{\avg{A|\mu}}$ is reported as a dashed line in
Fig. \ref{AmuFX}. The evidence for the law $|\avg{A|\mu}|\sim
1/\tau^\mu$ is very weak and hardly emerges from the noise level. In other
words, one cannot rule out the hypothesis that $\avg{A|\mu}=0$ and
that what is seen is just fluctuations due to finite sample
size. Furthermore there is no evidence of saturation for rare
events. This is consistent with the fact that FX markets are very
efficient and hence one expects a fairly large value of
$\chi$. 

The present statistical resolution does not allow for sharp statements
on the inefficiency of FX-markets. It serves however to illustrate the
practical issues that such an empirical analysis raises.

\appendix
\section*{Appendix}
\label{appAmu}

We give here some results of the algebraic manipulation of the model
defined in section~\ref{minor}, and the final equation, to be solved
numerically. In order to study the minima of $H$, as usual one
introduces a fictitious temperature $\beta$ and builds the partition
function $Z(\beta)=\Tr_m e^{-\beta H}$. In order to deal with quenched
disorder (i.e. with the heterogeneity of agents' trading strategies)
we resort to the replica trick \cite{MPV}. This amounts in averaging
the parition function $Z^n(\beta)$ of $n$ copies of the system over
the disorder and then taking the limit $n\to\infty$. This allows one
to compute the disorder average of $\log Z(\beta)$ which is a
self-averaging quantity. It is easy to see \cite{CMZ} that the
resulting free-energy is correctly described by a simple 
replica symmetric (RS) ansatz, i.e. 
\[
\frac{1}{N}\avg{\log Z(\beta)}\simeq -\beta f
\]
where
\[
f=\frac{\alpha}{2 \beta} 
\Bigg[  \ln \big(1+\frac{\beta \gamma}{ \alpha (\gamma + 2)} 
 (Q- q) \big) + \frac{\beta (1+q)}
{\alpha (\gamma + 2)/\gamma +\beta(Q- q)}\Bigg]\]
\[ +
\frac{\alpha}{2 \beta} \Big[QR-qr\Big] -\frac{1}{\beta}
\Bigg\lmoustache_0^1 \mbox{d}f\, \gamma\, f^{\gamma-1}\, \Bigg< \ln
Tr_m \exp\Big(-\beta f V_z(m|f)\Big)\Bigg>_z,
\]
where
\beas
Q&=&\sum_{\i=1}^N\,f_i^2\,m_i^2/\sum_{\i=1}^N\,f_i^2\\
q&=&\sum_{\i=1}^N\,f_i^2\,m_i m_i^{\prime}/\sum_{\i=1}^N\,f_i^2\\
V_{z}(m|f)&=& -\frac{\gamma+2}{\gamma}\frac{\alpha \beta f}{2}(R- r)
m^2 - \sqrt{(\gamma+2)\alpha r/\gamma}z m.
\eeas
Here $m_i$ and $m_i^\prime$ refer to two distinct replicas of the
system. Finally $R$ and $r$ arise as Langrange multipliers. The
analysis of the saddle point equations follows the usual steps (see
e.g. the appendics of Refs. \cite{MCZ,MMM}). The limit $\beta\to 0$
has to be taken in the end. The solution can be expressed in
parametric form in terms of a variable $\zeta$:

\beas
Q&=&q\\
Q&=&1-(\gamma+2)\int_0^1\!df f^{\gamma+1}\left[
\sqrt{\frac{2}{\pi}}\frac{e^{-f^2\zeta^2/2}}{f\zeta}+\left(1-\frac{1}{f^2
\zeta^2}\right){\rm erf}\left(\frac{f\zeta}{\sqrt{2}}\right)\right]
\eeas
and
\[
1+Q=\frac{\gamma+2}{\gamma}\frac{\alpha}{\zeta^2}.
\]

The critical threshold $\alpha_c$ is as usual
\cite{CMZ,MCZ,MMM,errata} derived imposing that the fraction of
non-frozen agents is qual to $\alpha$:
\[
\int_0^1[1-\phi(f)]\gamma f^{\gamma-1}df=\int_0^1{\rm
erf}\left(\frac{f\zeta}{\sqrt{2}}\right) \gamma f^{\gamma-1}df =\alpha.
\]
These equations can be easily solved numerically.


The analytic solution of the MG with generic distribution $\rho^\mu$
is described in Ref. \cite{relevance_mem} and follows the same lines
as above.  The parameters $\chi=\beta(Q-q)/\alpha$ and $Q$ are given
by the solution of the set of equations:

\beas
Q&=&1-\sqrt{\frac{2}{\pi}}\frac{e^{-\zeta^2/2}}{\zeta}-
\pr{1-\frac{1}{\zeta^2}}\erf\pr{\frac{\zeta}{\sqrt{2}}}\\
\alpha(1+Q)\avg{\frac{\tau^2}{[1+\tau\chi]^2}}_\tau&=&
\cro{\frac{\erf (\zeta/\sqrt{2})}{\chi\zeta}}^2\\
\Avg{\frac{\tau\chi}{1+\tau\chi}}_\tau&=&\frac{\erf(\zeta/\sqrt{2})}{\alpha}
\eeas

\noindent
where $\avg{\ldots}_\tau=\int_0^\infty\!d\tau\theta(\tau)\ldots$. 
These equations are valid in the symmetric phase
($\alpha>\alpha_c$). $\chi$ diverges at the critial point, where
$\alpha_c=\erf(\zeta/\sqrt{2})$.


\begin{thebibliography}{99}
\bibitem{notamu} If $x(t)$ is the FX rate at time $t$, we define
$\chi(t)=1$ if $x(t)>1.0003 x(t-1)$ and $\chi(t)=0$ otherwise. The event
$E^\mu(t)=\{Y(t)=\mu\}$ is defined in terms of the random variable

$$
Y(t)=\sum_{t'=t-M+1}^t \chi(t')2^{t-t'}.
$$
Then we measure conditional averages
$x^\mu=\langle x(t+1)-x(t)|Y(t)=\mu\rangle$ and 
$\sigma^\mu=\sqrt{\langle [x(t+1)-x(t)]^2|Y(t)=\mu\rangle}$. 
Finally we compute $A^\mu=x^\mu/\sigma^\mu$ and the frequency 
$f^\mu=\langle \delta_{Y(t),\mu} \rangle$.
\end{thebibliography}
\end{document}